\documentclass[sigconf,screen]{acmart}

\AtBeginDocument{%
  \providecommand\BibTeX{{%
    \normalfont B\kern-0.5em{\scshape i\kern-0.25em b}\kern-0.8em\TeX}}}

\setcopyright{acmlicensed}
\acmPrice{15.00}
\acmDOI{10.1145/3611643.3616256}
\acmYear{2023}
\copyrightyear{2023}
\acmSubmissionID{fse23main-p131-p}
\acmISBN{979-8-4007-0327-0/23/12}
\acmConference[ESEC/FSE '23]{Proceedings of the 31st ACM Joint European Software Engineering Conference and Symposium on the Foundations of Software Engineering}{December 3--9, 2023}{San Francisco, CA, USA}
\acmBooktitle{Proceedings of the 31st ACM Joint European Software Engineering Conference and Symposium on the Foundations of Software Engineering (ESEC/FSE '23), December 3--9, 2023, San Francisco, CA, USA}
\received{2023-02-02}
\received[accepted]{2023-07-27}

\usepackage{adjustbox}
\usepackage{multirow}
\usepackage{graphicx}
\usepackage{arydshln}
\usepackage{hyperref}
\usepackage[capitalize, nameinlink]{cleveref}

\begin{document}
\title{RAP-Gen: Retrieval-Augmented Patch Generation with CodeT5 for Automatic Program Repair}

\author{Weishi Wang}
\authornote{Equal contribution. Corresponding author: wang.y@salesforce.com.}
\email{weishi001@e.ntu.edu.sg}
\affiliation{
  \institution{Salesforce AI Research}
  \institution{Nanyang Technological University}
  \country{Singapore}
}
\author{Yue Wang}
\authornotemark[1]
\email{wang.y@salesforce.com}
\affiliation{
  \institution{Salesforce AI Research}
  \country{Singapore}
}

\author{Shafiq Joty}
\email{sjoty@salesforce.com}
\affiliation{
  \institution{Salesforce AI Research}
  \country{USA}
}

\author{Steven C.H. Hoi}
\email{shoi@salesforce.com}
\affiliation{
  \institution{Salesforce AI Research}
  \country{Singapore}
}

\begin{abstract}
Automatic program repair (APR) is crucial to reduce manual debugging efforts for developers and improve software reliability. While conventional search-based techniques typically rely on heuristic rules or a redundancy assumption to mine fix patterns, recent years have witnessed the surge of deep learning (DL) based approaches to automate the program repair process in a data-driven manner. However, their performance is often limited by a fixed set of parameters to  model the highly complex search space of APR. 

To ease such burden on the parametric models, in this work, we propose a novel \textbf{R}etrieval-\textbf{A}ugmented \textbf{P}atch \textbf{Gen}eration framework (RAP-Gen) by explicitly leveraging relevant fix patterns retrieved from a codebase of previous bug-fix pairs. Specifically, we build a hybrid patch retriever to account for both lexical and semantic matching based on the raw source code in a language-agnostic manner, which does not rely on any code-specific features. In addition, we adapt a code-aware language model CodeT5 as our foundation model to facilitate both patch retrieval and generation tasks in a unified manner. We adopt a stage-wise approach where the patch retriever first retrieves a relevant external bug-fix pair to augment the  buggy  input for the CodeT5 patch generator, which synthesizes a ranked list of repair patch candidates. Notably, RAP-Gen is a generic APR framework that can flexibly integrate different patch retrievers and generators to repair various types of bugs.

We thoroughly evaluate RAP-Gen on three benchmarks in two programming languages, including the TFix benchmark in JavaScript, and Code Refinement and Defects4J benchmarks in Java, where the bug localization information may or may not be provided. Experimental results show that RAP-Gen significantly outperforms previous state-of-the-art (SoTA) approaches on all benchmarks, e.g., boosting the accuracy of T5-large on TFix from 49.70\% to 54.15\% (repairing 478 more bugs) and repairing 15 more bugs on 818 Defects4J bugs. Further analysis reveals that our patch retriever can search for relevant fix patterns to guide the APR systems. 
\end{abstract}



\maketitle

\section{INTRODUCTION}
Program repair is one of the most important stages to maintain software quality, which however is a  time-consuming and cost-dominating process in modern software development~\cite{DBLP:conf/msr/WeissPZZ07,planning2002economic}.
Therefore, there have been huge needs for Automatic Program Repair (APR) tools to ease the difficulty and cost of program repair for developers with use cases including search of patches at program development time \cite{DBLP:conf/oopsla/MusluBHEN12}, build time~\cite{Urli2017HowTD,Martin2019RepairnatorPP} or run time~\cite{Perkins2009AutomaticallyPE,Durieux2017ProductionDrivenPG}. 

A notable class of conventional techniques for APR is known
as search-based (also referred to as generate-and-validate) approach~\cite{DBLP:journals/tse/GouesNFW12,Weimer2009AutomaticallyFP,DBLP:conf/icse/QiMLDW14, DBLP:conf/icse/WenCWHC18,DBLP:conf/wcre/LiuZ18,DBLP:conf/issta/JiangXZGC18}.
They often search for repairs based on the \emph{fix patterns} mined via manual heuristic rules~\cite{DBLP:conf/icse/KimNSK13,DBLP:conf/icse/QiMLDW14,tan2015relifix} or redundancy-based techniques~\cite{DBLP:journals/tse/GouesNFW12,Long2016AutomaticPG,DBLP:conf/sigsoft/LongR15, DBLP:conf/icse/WenCWHC18,DBLP:conf/wcre/LiuZ18,DBLP:conf/issta/JiangXZGC18}.
The latter group of approaches make a \emph{redundancy assumption}~\cite{White2019SortingAT}  that
the fixed patch can often be found (or reconstructed) from elsewhere in the codebase (a donor code snippet). This hypothesis has been validated empirically by studies~\cite{DBLP:conf/sigsoft/BarrBDHS14,Martinez2014DoTF} showing that a significant proportion of commits (3\%-17\%) are indeed composed of existing codebase.

Meanwhile, with the recent advancement in deep learning technologies, numerous deep learning (DL)-based  APR approaches~\cite{DBLP:journals/tosem/TufanoWBPWP19,DBLP:journals/tse/ChenKTPPM21,DBLP:conf/issta/LutellierPPLW020,DBLP:conf/icse/JiangL021,DBLP:conf/sigsoft/ZhuSXZY0Z21,DBLP:conf/icml/BerabiHRV21} have been proposed to automate the   repair process via  parametric models in a purely data-driven manner. 
In this paradigm, the APR task is typically formulated as a neural machine translation (or sequence-to-sequence learning) problem~\cite{sutskever2014sequence} in order to translate a buggy (source) program into a correct (target) version.
Despite their promising results in software intelligence tasks, their performance is often limited by the fixed set of model parameters to learn the highly complex distributional patterns for program repair, even with several hundreds of million parameters~\cite{DBLP:conf/icse/JiangL021,DBLP:conf/sigsoft/ZhuSXZY0Z21,DBLP:conf/icml/BerabiHRV21}.

\begin{figure}[!t]
  \centering
  \includegraphics[width=1\linewidth]{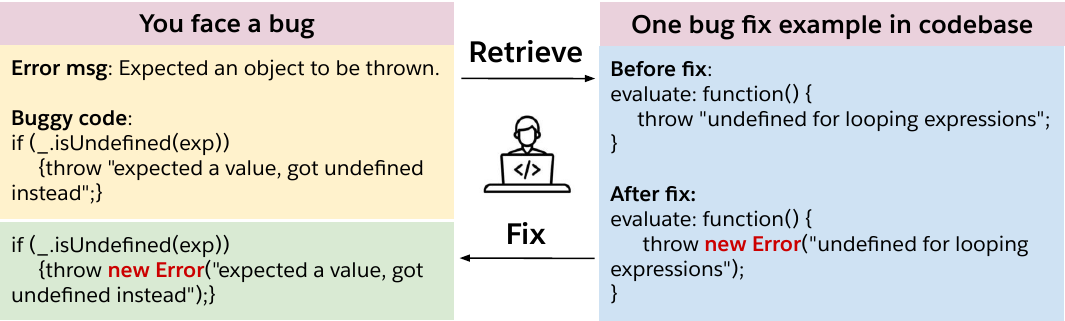}
  \caption{One motivating example of how developers fix a bug by referring to a retrieved  fix pattern in codebase.}
  \label{fig:motivate}
  \vspace{-1em}
\end{figure}

To ease such burden on the parametric neural models, in this work, we propose a novel  retrieval-augmented patch generation framework called RAP-Gen to additionally leverage relevant fix patterns from a patch retriever.
Earlier APR techniques based on the redundancy assumption have shown that mining fix patterns from existing codebase~\cite{DBLP:journals/tse/GouesNFW12,DBLP:conf/issta/JiangXZGC18} or even external Q\&As from StackOverflow~\cite{DBLP:conf/wcre/LiuZ18} can serve as crucial repair ingredients for APR.
Our model, which is semi-parametric in nature, aims to combine both benefits of  the implicit (parametric) end-to-end program repair learning and the explicit (non-parametric) fix pattern mining.
One distinction from prior fix pattern mining work is that we utilize the top relevant \emph{bug-fix pair} as a guiding fix pattern for a buggy patch instead of clustering the fix templates with hand-crafted heuristics.
This retrieval-guiding strategy is also motivated by debugging behaviours of program  developers, who often search for relevant bug-fix examples to distill some repair clues for bug fixing. 
\cref{fig:motivate} illustrates a motivating example, where we can find that  the retrieved previous repair example informs a fix pattern of wrapping the string with an ``Error'' object for the ``throw'' statement, which guides the developer to fix the target bug under consideration.

In addition, we propose to adapt a Transformer-based~\cite{DBLP:conf/nips/VaswaniSPUJGKP17}  encoder-decoder model CodeT5~\cite{DBLP:conf/emnlp/0034WJH21} as the unified foundation model of RAP-Gen for both patch retrieval and generation tasks.
CodeT5 is a generic code-aware language model pretrained on large source code corpora in eight popular programming languages (including JavaScript and Java) curated from GitHub, achieving state-of-the-art (SoTA) performance in both code understanding and generation tasks.
RAP-Gen adopts a stage-wise learning strategy to connect the patch retriever and  patch generator:
the patch retriever first searches for a relevant bug fix pattern and then pass it to the CodeT5 patch generator to synthesize a ranked list of fix patch candidates based on both the source buggy code and the retrieved external bug fix knowledge.
While such retrieval-augmented generation paradigm has been explored in other tasks such as question answering~\cite{DBLP:conf/emnlp/KarpukhinOMLWEC20} and code generation and summarization~\cite{DBLP:conf/emnlp/ParvezACRC21}, we are the first to investigate its effectiveness for  APR systems based on large-scale pretrained language models for code. 

For the retrievers, we propose a hybrid approach that accounts for both lexical and semantic matching through sparse (BM25~\cite{DBLP:journals/ftir/RobertsonZ09}) and dense (DPR~\cite{DBLP:conf/emnlp/KarpukhinOMLWEC20}) retrieval based on the raw source code. 
We employ CodeT5's encoder as our dense DPR retriever and propose to train it  with a contrastive learning objective~\cite{DBLP:journals/corr/abs-1807-03748} using previous bug-fix pairs as the fix patch often shares most of semantics with its buggy patch.
The dense DPR retriever is expected to capture  deeper code semantics  while the sparse keyword-based BM25 retriever focuses more on the lexical similarity which is sensitive to the choice of naming for code identifiers.
Notably, the hybrid retriever is language-agnostic as it does not require any code-specific features such as abstract syntax trees (ASTs). Experiments reveal that our patch retriever is able to retrieve lexically and semantically relevant fix patterns to guide APR systems.

We investigate the effectiveness of RAP-Gen  in different APR scenarios including JavaScript linter-raised diagnostics (TFix~\cite{DBLP:conf/icml/BerabiHRV21}), Java bug-fix commits (Code Refinement~\cite{DBLP:journals/tosem/TufanoWBPWP19}), and real Java bugs accompanied with test cases in  open source projects (Defects4J~\cite{DBLP:conf/issta/JustJE14}).
Among these benchmarks, we formulate the APR problem as that given a buggy code patch, the APR model learns to predict a fix patch that repairs the bug from a codebase of previous bug-fix pairs written by developers.
The correctness of the predicted fix patches are validated against either static analyzers (TFix) or unit testing (Defects4J), or via a direct comparison with the ground-truth fixes written by developers.
Overall, extensive experimental results show that our RAP-Gen significantly outperforms existing DL-based methods on all these three APR benchmarks.

In summary, the paper makes the following contributions:
\begin{itemize}
\vspace{-0.5em}
\itemsep0em
\item We propose a novel retrieval-augmented patch generation framework (RAP-Gen) for APR. It is a generic framework that can be easily integrated with any sequence-to-sequence learning models. To the best of our knowledge, this is the first work to leverage the power of retrieval in fix pattern mining for DL-based APR systems.

\item We present a hybrid patch retriever for fix pattern mining that accounts for both lexical and semantic matching through a combination of sparse and dense retrievers. It is a language-agnostic patch retriever  using raw source code which does not require any code-specific features. 

\item We propose to adapt a generic pretrained code-aware language model CodeT5 as a foundation model for RAP-Gen to fix various bugs. Moreover, we leverage it for both patch retrieval and generation task in a unified manner.

\item We extensively evaluate RAP-Gen on three APR benchmarks in JavaScript and Java. Results show  RAP-Gen significantly outperforms SoTA DL-based methods on all benchmarks. Particularly, our best model yields substantial improvements (49.70 $\rightarrow$ 54.15 on exact match accuracy and 69.30 $\rightarrow$ 78.80 on error removal accuracy) on TFix over the previous SoTA T5-large model with a 3.5x larger model size than ours. On Code Refinement, RAP-Gen sets new SoTA exact match results of 24.80 and 15.84 over CodeT5's 21.61 and 13.96 for the small and medium subsets. 
On Defects4J, RAP-Gen achieves new SoTA performance, repairing 15 more bugs (110 $\rightarrow$ 125) with perfect FL  and 6 more bugs (68 $\rightarrow$ 74) without perfect FL than the previous SoTA models.

\end{itemize}

\begin{figure*}[htbp]
  \centering
  \includegraphics[width=0.83\linewidth]{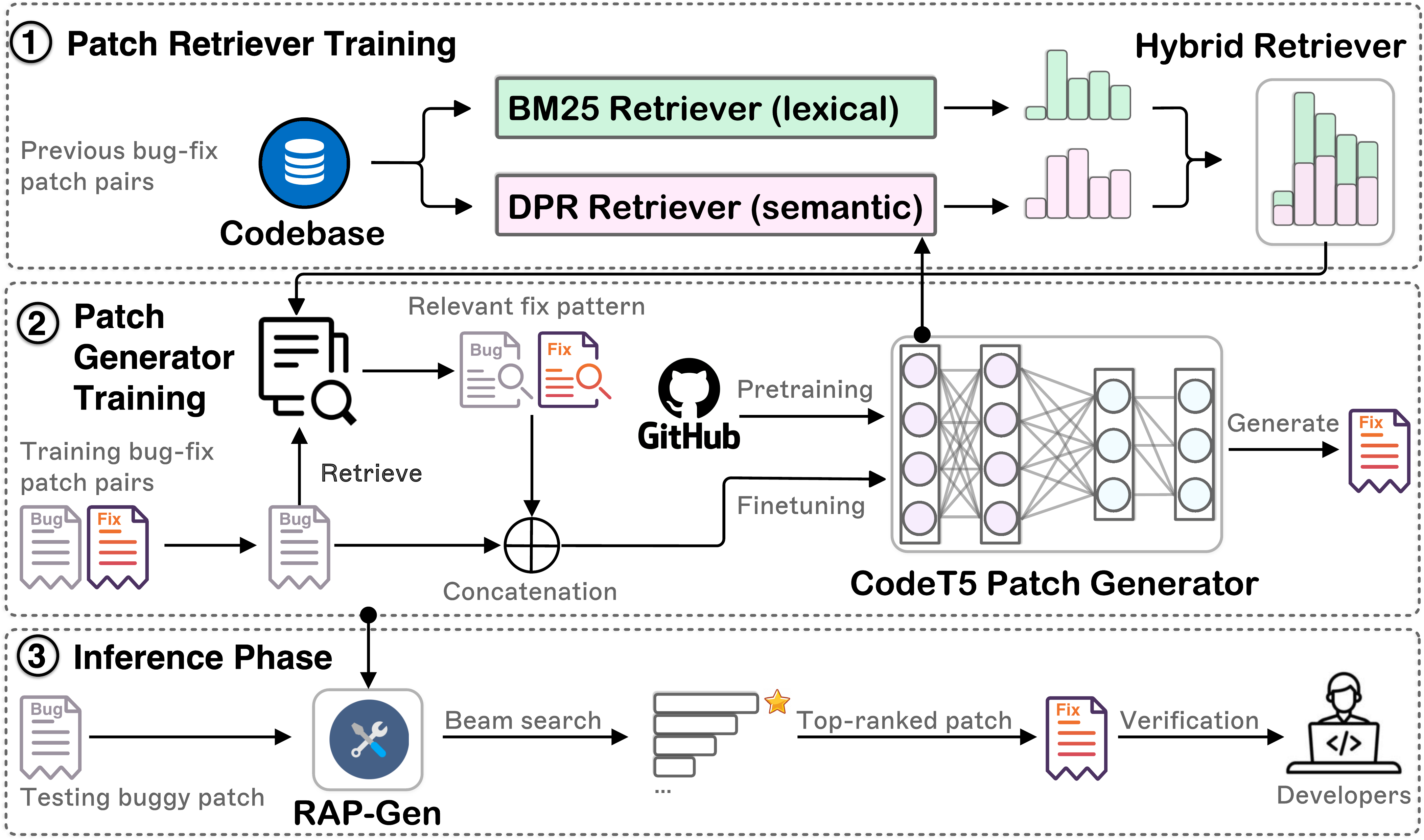}
  \vspace{-0.5em}
  \caption{Retrieval-augmented patch generation (RAP-Gen) framework with CodeT5 for automatic program repair. We first retrieve the relevant bug fix patterns from the codebase through our hybrid patch retriever which takes both lexical and semantic similarity into account. We then concatenate the top-1 retrieved bug fix pattern with the query buggy patch for our patch generator to synthesize a ranked list of fix patch candidates for developers to verify.}
  \label{fig:racodet5_model}
  \vspace{-0.5em}
\end{figure*}
\section{RELATED WORKS}
\subsection{Automatic Program Repair}
In the past decades, automatic program repair (APR)  has attracted growing attention and various APR techniques have been proposed to reduce the manual efforts in debugging.
A notable class of conventional techniques for APR is known as search-based (or generate-and-validate) approach~\cite{DBLP:journals/tse/GouesNFW12,Weimer2009AutomaticallyFP,DBLP:conf/icse/QiMLDW14, DBLP:conf/icse/WenCWHC18,DBLP:conf/wcre/LiuZ18,DBLP:conf/issta/JiangXZGC18}.
Earlier search-based APR techniques are often based on program modification or mutation with  heuristic algorithm~\cite{DBLP:conf/icse/QiMLDW14} or genetic programming~\cite{DBLP:journals/tse/YuanB20} to produce a large pool of candidate fixes for validating with unit tests. 
The search strategy has been further extended to adopt \emph{fix patterns} mined using redundancy-based techniques~\cite{DBLP:journals/tse/GouesNFW12,Long2016AutomaticPG,DBLP:conf/sigsoft/LongR15, DBLP:conf/wcre/LeLG16,DBLP:conf/icse/WenCWHC18,DBLP:conf/wcre/LiuZ18,DBLP:conf/issta/JiangXZGC18}.
These  approaches make a \emph{redundancy assumption}~\cite{White2019SortingAT} that
the fixed patch can often be reconstructed from elsewhere in the codebase, which has been validated empirically by studies~\cite{DBLP:conf/sigsoft/BarrBDHS14,Martinez2014DoTF} showing that a significant proportion of commits (3\%-17\%) are indeed composed of existing codebase.
More redundancy-based techniques have shown that mining fix patterns from existing codebase~\cite{DBLP:journals/tse/GouesNFW12,DBLP:conf/issta/JiangXZGC18} or even external Q\&As from StackOverflow~\cite{DBLP:conf/wcre/LiuZ18} can largely benefit APR systems.

Recently, with the recent advancement in deep learning (DL) approaches for natural language processing (NLP), many DL-based APR techniques~\cite{DBLP:journals/tosem/TufanoWBPWP19,DBLP:journals/tse/ChenKTPPM21,DBLP:conf/issta/LutellierPPLW020,DBLP:conf/icse/JiangL021,DBLP:conf/sigsoft/ZhuSXZY0Z21,DBLP:conf/icml/BerabiHRV21,DBLP:conf/emnlp/Bui0H22} have been proposed to automate the program repair process in an end-to-end data-driven manner. 
Motivated by the success of Neural Machine Translation (NMT), these techniques often formulate APR as a sequence-to-sequence NMT problem~\cite{sutskever2014sequence}, which is to translate a buggy program into a fixed  version. 
Various neural architectures have been explored in learning-based  APR techniques.
Earlier techniques~\cite{DBLP:journals/tosem/TufanoWBPWP19, DBLP:journals/tse/ChenKTPPM21} are based on recurrent neural networks~\cite{hochreiter1997long}, which is further extended to convolution neural networks~\cite{DBLP:conf/icml/GehringAGYD17} in CoCoNuT~\cite{DBLP:conf/issta/LutellierPPLW020} and Transformer-based models~\cite{DBLP:conf/nips/VaswaniSPUJGKP17} by many recent DL-based models including TFix~\cite{DBLP:conf/icml/BerabiHRV21}, CURE~\cite{DBLP:conf/icse/JiangL021}, Recoder~\cite{DBLP:conf/sigsoft/ZhuSXZY0Z21}, RewardRepair~\cite{DBLP:conf/icse/YeMM22},
and SelfAPR~\cite{SelfAPR}. 
Notably, many of these DL-based approaches explore improving APR by leveraging code-specific features such as abstract syntax trees (ASTs)~\cite{DBLP:conf/sigsoft/ZhuSXZY0Z21,DBLP:conf/icse/Li0N22} and test execution diagnostics~\cite{DBLP:conf/icse/YeMM22,SelfAPR}.
Specifically, Recoder~\cite{DBLP:conf/sigsoft/ZhuSXZY0Z21} learns the syntax-guided edits over the ASTs to ensure the syntactic correctness of the generated fix patch, while DEAR~\cite{DBLP:conf/icse/Li0N22} uses tree-based Long Short-Term Memory (LSTM) model~\cite{DBLP:conf/acl/TaiSM15} to better encode the code structure and constructs a suitable fixing context using  surrounding AST subtrees.
For the use of test execution information, SelfAPR~\cite{SelfAPR} encodes test execution diagnostics into the input representation,
while RewardRepair~\cite{DBLP:conf/icse/YeMM22} improves APR with a loss function based on both program compilation and test execution information.

In terms of APR benchmarks, the most popular one would be Defects4J~\cite{DBLP:conf/issta/JustJE14}, which contains real bug-fix patches from open source GitHub projects and has been widely adopted by a large body of APR  work~\cite{DBLP:conf/issta/LutellierPPLW020,DBLP:conf/icse/JiangL021,DBLP:conf/sigsoft/ZhuSXZY0Z21,DBLP:conf/icse/YeMM22,SelfAPR}. One notable feature of this benchmark is that it contains a test suite to  validate whether the bugs are fixed or not.
However, as these APR approaches rely on test cases, they are inapplicable to newly discovered bugs or bugs difficult to test for deterministically~\cite{DBLP:conf/icse/TonderG18}. 
Additionally, it remains a key challenge to obtain a large-scale APR dataset with test cases, e.g., one of the largest one Defects4J only contains less than 1000 bugs and another popular one QuixBugs~\cite{QuixBugs} only have 40 bugs.
To get rid of the requirement of test cases, there is another group of APR research~\cite{DBLP:conf/icse/TonderG18,Phoenix,SpongeBugs,PyTER, DBLP:conf/icml/BerabiHRV21} focusing on static analysis bugs or violations, which can be  flagged by static analysis tools and is easier to curate  much more bug-fix data. 
Besides, another type of APR~\cite{DBLP:journals/tosem/TufanoWBPWP19} is based on the bug-fixing commits by checking whether the commit comments contain some keywords such as ``repair'' and ``fix''. We consider all these types of APR use cases in this work.

\subsection{Pretrained Language Models for Code}
Pretrained language models (LMs) like  GPT~\cite{radford2018improving}, BERT~\cite{DBLP:conf/naacl/DevlinCLT19}, and T5~\cite{DBLP:journals/jmlr/RaffelSRLNMZLL20}  have significantly boosted  performance  in a broad set of NLP tasks. Inspired by their success, much recent work~\cite{DBLP:conf/emnlp/FengGTDFGS0LJZ20,DBLP:conf/iclr/GuoRLFT0ZDSFTDC21,DBLP:journals/corr/abs-2102-04664,DBLP:journals/corr/abs-2105-08645,DBLP:conf/naacl/AhmadCRC21,DBLP:journals/corr/abs-2107-03374,DBLP:conf/emnlp/0034WJH21} attempts to adapt the NLP pretraining methods to programming language. They often rely on  either an encoder-only BERT-style models  (CodeBERT~\cite{DBLP:conf/emnlp/FengGTDFGS0LJZ20} and GraphCodeBERT~\cite{DBLP:conf/iclr/GuoRLFT0ZDSFTDC21}) or decoder-only GPT-style models (CodeGPT~\cite{DBLP:journals/corr/abs-2102-04664} and Codex~\cite{DBLP:journals/corr/abs-2107-03374}), or encoder-decoder models (PLBART~\cite{DBLP:conf/naacl/AhmadCRC21} and CodeT5~\cite{DBLP:conf/emnlp/0034WJH21,wang2023codet5plus,DBLP:conf/nips/Le0GSH22}).
Particularly, CodeT5 is a unified  language model pretrained with a code-aware pretraining objective on large-scale code corpora covering 8 different programming languages, which has been shown to achieve SoTA performance on a wide range of code understanding and generation tasks~\cite{DBLP:journals/corr/abs-2102-04664}.
Compared to previous DL-based APR approaches such as CURE~\cite{DBLP:conf/icse/JiangL021} and TFix~\cite{DBLP:conf/icml/BerabiHRV21} that utilize LMs pretrained  primarily on natural language corpus,  we propose to leverage the code-aware LMs of CodeT5~\cite{DBLP:conf/emnlp/0034WJH21} for  APR with better code understanding capability.

There are recent attempts~\cite{DBLP:journals/corr/abs-2208-11640,DBLP:conf/icse-apr/PrennerBR22} to explore few-shot learning of large language models (LLMs) for APR. According to \citet{DBLP:conf/icse-apr/PrennerBR22}, their method based on Codex achieves 46\% EM compared to the finetuned T5’s 59\% on a random sample of 200 instances from TFix, showing that there is still a gap between few-shot learning and finetuning results. 
Besides, few-shot learning of LLMs requires more engineering efforts for prompting tuning and post-processing~\cite{DBLP:journals/corr/abs-2208-11640}, which is labor-intensive. Another concern is that LLMs such as Codex~\cite{DBLP:journals/corr/abs-2107-03374} are not open sourced and it might be expensive to use their APIs, e.g., the Davinci version costs \$0.02 for every 1K tokens\footnote{\url{https://openai.com/api/pricing/}}.

\subsection{Retrieval-Augmented Generation}
A general retrieval-augmented generation paradigm is comprised of three components including information retrieval, data augmentation and generation model~\cite{DBLP:journals/corr/abs-2202-01110}. 
It has been widely studied in NLP and shown to achieve SoTA performance in a wide range of NLP tasks including question answering and question generation~\cite{DBLP:conf/nips/LewisPPPKGKLYR020,DBLP:conf/eacl/IzacardG21} and machine translation~\cite{DBLP:conf/aaai/GuWCL18}.
Inspired by their success, much research work adapts this paradigm (also referred as retrieve-and-edit/refine framework) to benefit software intelligence tasks, including code autocompletion~\cite{DBLP:conf/nips/HashimotoGOL18,DBLP:journals/corr/abs-2203-07722}, code summarization~\cite{DBLP:conf/emnlp/ParvezACRC21,DBLP:conf/kbse/LiL000J21}, and code generation~\cite{DBLP:conf/emnlp/ParvezACRC21,wang2023codet5plus}.

\section{APPROACH}
We propose RAP-Gen, a novel retrieval-augmented patch generation
framework for APR, which aims to improve APR performance by leveraging a relevant bug fix pattern  retrieved from a codebase of previous bug-fix pairs. 
As shown in~\cref{fig:racodet5_model}, our RAP-Gen framework consists of three stages: 1) a patch retriever training stage to learn a hybrid retriever that can find relevant code patches based on the lexical and semantical similarity; 2) a patch generator training stage to train a CodeT5 model to produce the fix patch based on both buggy input and retrieved bug-fix examples; 3) an inference stage to predict multiple fix patches where the top-ranked one will be passed to developers for verification. 

Note that while retrieval-augmented generation techniques have been explored in many NLP tasks~\cite{DBLP:conf/nips/LewisPPPKGKLYR020,DBLP:conf/eacl/IzacardG21}, it is not trivial to adapt such techniques to APR tasks and requires systematic adaptation to address some unique challenges.
 The first challenge is how to retrieve relevant fix patterns for effectively guiding APR, where we build a hybrid retriever based on both lexical and semantic information, which is analyzed and compared with other retrievers in Table~\ref{table:retriever_eval}.
The second challenge is how to build a top-performing APR model for various languages and APR scenarios. We leverage a language-agnostic pretrained model CodeT5 for both retrieval and patch generation, which is a more unified approach compared to prior work~\cite{DBLP:conf/emnlp/ParvezACRC21,DBLP:journals/corr/abs-2203-07722} requiring a different retriever and generator. 

In the remainder of this section, we first introduce the task formulation of the retrieval-augmented patch generation for APR in \cref{model:task} and then revisit  the backbone model of CodeT5 in \cref{ssec:codet5}, followed by detailing  the \emph{hybrid patch retriever} in \cref{model:retriever} and the \emph{retrieval-augmented patch generator} in \cref{model:generator}.


\subsection{Task Formulation}\label{model:task}
Let $\mathcal{D} = \{(X_i,Y_i)\}_{i=1}^{|\mathcal{D}|}$ be a program repair dataset consisting of $|\mathcal{D}|$ bug-fix pairs $(X_i,Y_i)$, where $X_i$ and $Y_i$ are the $i$-th buggy and fixed program patch, respectively.
Assume that we have a codebase containing a large collection of previous bug-fix pairs $\mathcal{C} = \{(B_j,F_j)\}_{j=1}^{|\mathcal{C}|}$, where  $(B_j,F_j)$ denotes the $j$-th previous bug-fix pair.
Given a  buggy program patch $X_i$ in $\mathcal{D}$, a retriever retrieves  the most relevant bug-fix pair  $(B_j,F_j)$ in the codebase $\mathcal{C}$  based on a relevance scoring function $f_\phi(X_{i},B_{j})$ parameterized by $\phi$. 
Then the original input sequence $X_i$ is augmented with the retrieved bug-fix pair to form a new input sequence $\hat{X}_i = X_i \oplus B_j \oplus F_j$, where $\oplus$ denotes the concatenation operation. The sequence-to-sequence (seq2seq) generator then generates $Y_i$ from $\hat{X}_i$ in an autoregressive manner. Formally, we aim to learn the following probability with the patch seq2seq generator parameterized by $\theta$:
\begin{displaymath}
P_{\theta}(Y_i|\hat{X}_i)=\prod_{k=1}^{n}P_{\theta}
(Y_{i,k}|\hat{X}_i,Y_{i,1}:Y_{i,k-1}) 
\end{displaymath}
where $Y_{i,1}:Y_{i,k-1}$ is the previous sequence before the $k$-th token and $n$ denotes the number of tokens in the target sequence $Y_i$. 
Note that we regard the external codebase $\mathcal{C}$ as a non-parametric memory and the retrieved bug-fix pair as a guiding fix pattern for the generator.
In probabilistic terms, the retrieval  can be formulated as a latent variable $Z_j = (B_j, F_j)$, which is approximated by top-1 in our case. Formally, the probability can be decomposed as:
\begin{displaymath}
\small 
  P(Y_i|X_i) = \sum_{j=1}^{|\mathcal{C}|} \underbrace{P_{\phi}(Z_j|X_i)}_{\text{Retriever}} \underbrace{P_{\theta}(Y_i|X_i, Z_j)}_{\text{Generator}}  \approx   P_{\theta}(Y_i|X_i, Z_j^{*})  
 \normalsize
\end{displaymath}
\noindent where $Z_j^{*}$ is the top-1 retrieved output from the retriever $P_{\phi}(Z_j|X_i)$.  
We adopt such top-1 approximation as marginalization over large $k$ makes the training and inference complicated and inefficient~\cite{DBLP:conf/nips/LewisPPPKGKLYR020}. We also tried to employ top-$k$ ($k=2,3,5$) with the Fusion-in-Decoding or FiD method~\cite{DBLP:conf/eacl/IzacardG21} but did not observe a salient performance improvement.

\subsection{Revisiting CodeT5}\label{ssec:codet5}
CodeT5~\cite{DBLP:conf/emnlp/0034WJH21} is a unified pretrained Transformer-based encoder-decoder language model that achieves SoTA results in both code understanding and generation tasks. It is pretrained on 8.3 million functions in 8 different programming languages (i.e., Ruby, JavaScript, Go, Python,
Java, PHP, C, C\#) collected from GitHub. CodeT5 employs a set of identifier-aware pretraining objectives to incorporate the code-specific knowledge into the language model. In this work, we adapt CodeT5 as our dense DPR retriever and patch generator to harness its powerful code understanding capability.

\paragraph{\textbf{BPE Subword Tokenization}}
One benefit of using CodeT5 is that it provides a code-specific Byte-Pair Encoding (BPE)~\cite{DBLP:conf/acl/SennrichHB16a} tokenizer.
It can avoid the prevalent Out-of-Vocabulary (OoV) problems in the code domain as programmers tend to write arbitrary identifiers~\cite{DBLP:conf/ijcai/LiWLK18} and it is impossible to build a fixed vocabulary to accommodate arbitrary tokens (commonly known as \emph{open vocabulary} problem \cite{DBLP:conf/acl/SennrichHB16a}). BPE is an algorithm that learns how to efficiently split tokens into subwords based on their frequency distribution. It can also help reduce the vocabulary size as it will split rare tokens into multiple subwords instead of directly adding the whole tokens into the vocabulary. 
Additionally, as the CodeT5 tokenizer is pretrained and optimized for eight popular programming languages, the resulting tokenization generalizes well. As pointed out by~\cite{DBLP:conf/emnlp/0034WJH21}, it reduces the tokenized sequence by 30\% - 45\% on average compared to the default T5 tokenizer~\cite{DBLP:journals/jmlr/RaffelSRLNMZLL20}. 

\paragraph{\textbf{Encoder and Decoder Architecture}}
CodeT5 consists of a stack of  Transformer layers~\cite{DBLP:conf/nips/VaswaniSPUJGKP17} for its encoder and decoder. Each Transformer layer contains a multi-head self-attention for feature aggregation  followed by a feed forward layer over the output of previous layer. The final layer produces the hidden states for all input tokens, which can be employed as the code presentation for classification or generation tasks. For the CodeT5 encoder, it utilizes bidirectional attention masks to learn better contextualized representation similar to BERT~\cite{DBLP:conf/naacl/DevlinCLT19}, while  the CodeT5 decoder employs causal attention masks to ensure each token can only attend to the previous tokens for better sequence generation. In RAP-Gen framework, we adapt the CodeT5 as the patch generator and its encoder specifically for the dense retriever. 

\subsection{Hybrid Patch Retriever}\label{model:retriever}
The retriever module in  RAP-Gen  aims to retrieve relevant fix patterns to guide the APR process. 
It builds on a relevance scoring function $f_{\phi}(X_{i},B_{j})$ to compute the relevance between the (query) bug $X_i$ in $\mathcal{D}$ and a previous (key) bug $B_j$ in the codebase $\mathcal{C}$.  As shown in ~\ref{fig:racodet5_model}~{1}, we utilize a hybrid approach to combine  a lexical-based   BM25~\cite{DBLP:journals/ftir/RobertsonZ09} retriever and a semantic-based  DPR~\cite{DBLP:conf/emnlp/KarpukhinOMLWEC20} retriever to take both lexical and semantic information into account. 
Prior work like~\cite{DBLP:conf/emnlp/KarpukhinOMLWEC20} show that sparse and dense retriever can complement each other for more robust text retrieval.

\paragraph{\textbf{Lexical-based Retriever}} 
We employ BM25~\cite{DBLP:journals/ftir/RobertsonZ09}, a well-known term-based retriever that uses sparse vector representation for lexical matching. BM25 converts each code patch as bag-of-words representation and computes a lexical similarity  between the query patch $X_i$ and a candidate patch $B_j$. The computed similarity score is represented as $f_{\phi}(X_{i},B_{j}) = BM25(X_i, B_j)$. As a sparse term-based retriever, BM25 is sensitive to the choice of identifier naming in source code which does not impact the code semantics.

\begin{figure*}[!ht]
  \centering
  \includegraphics[width=0.99\linewidth]{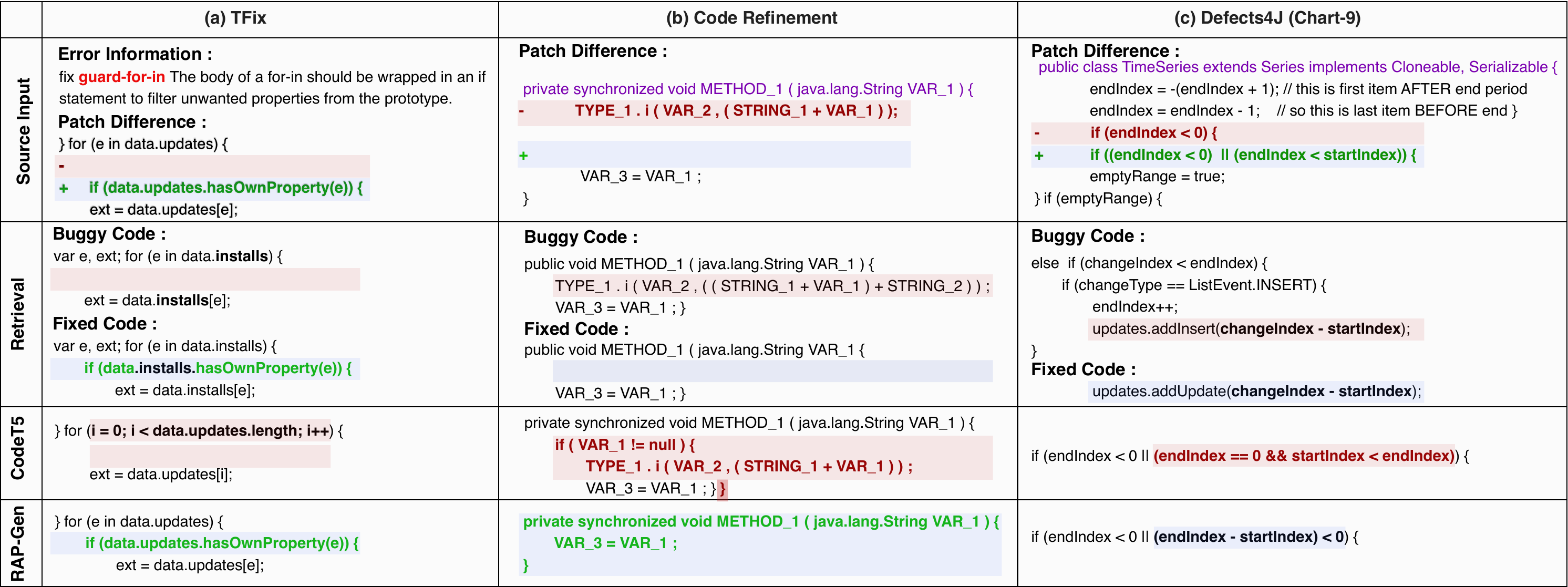}
  \caption{Bug fix examples  on three APR benchmarks, where RAP-Gen successfully fix bugs while CodeT5  fails to do so.}
  \label{fig:case_study}
\end{figure*}

\paragraph{\textbf{Semantic-based Retriever}}
We employ Dense Passage Retriever (DPR)~\cite{DBLP:conf/emnlp/KarpukhinOMLWEC20} to retrieve relevant patches via measuring their semantic similarity. 
To encode the code patch, we use a Transformer-based encoder to map each patch to a fixed-size dense vector. Specifically, we initialize  the DPR  from a pretrained CodeT5 encoder and train it for a code-to-code retrieval task.
For training the DPR, we propose to use the bug-fix pairs in the codebase by considering the buggy code $B_j$ as the query and the corresponding fixed code $F_j$ as the key. This is based on the assumption that the  buggy patch and its fixed patch  often shares similar semantics (e.g., identifiers and code structures). This trick avoids the massive manual annotation efforts needed to curate a bug-to-bug search dataset.

For each query patch and candidate patch, we prepend a special token of  \texttt{[CLS]} into its tokenized sequence and employ the final layer hidden state of the \texttt{[CLS]} token as the patch representation.
We use a shared DPR to separately encode the query patch  $X_i$ in $\mathcal{D}$ and a candidate patch $B_j$ in $\mathcal{C}$ as $CLS_{X_i}$ and $CLS_{B_j}$, respectively. 
Then the similarity is computed by the inner product between these two patch representations as the following:
\begin{displaymath}
  f_{\phi}(X_{i},B_{j}) = sim(X_i,B_j) = [CLS_{X_i}]^{T} [CLS_{B_j}]
\end{displaymath}

For training the DPR retriever, we leverage the \emph{in-batch negatives} to optimize an InfoNCE contrastive loss~\cite{DBLP:journals/corr/abs-1807-03748}  defined as follows:
\begin{displaymath}\label{eq:contrast}
\mathcal{L}_{nce}=\frac{1}{N} \sum_{i=1}^{N} -\log \frac{\exp (sim(B_{i},F_{i}))}{\exp (sim(B_{i},F_{i})) + \sum_{j \in \mathcal{M},j \neq i} \exp (sim(B_{i},F_{j}))}
\end{displaymath}
where $\mathcal{M}$ is the current minibatch and $N$ denotes the number of positive training examples in the minibatch. This objective aims to maximize the similarity between positive examples while minimizing the similarity between negative examples. Each positive example will have $|\mathcal{M}|-1$ negative samples.
Note that we do not adopt the hard negative mining strategy as in~\cite{DBLP:conf/emnlp/KarpukhinOMLWEC20} due to the noisy nature of the training data. 

In the inference stage, given a query buggy patch $X_i$, the DPR retrieves a relevant bug-fix pair $(B_j,F_j)$ by computing the similarity between $X_i$ (query) and $B_j$ (key). We also tried to base on the similarity between  $X_i$ and $F_j$ but  it did not yield better results.

\paragraph{\textbf{Hybrid Retriever}} 
To take both lexical and semantic information into account, we utilize a hybrid approach following~\cite{DBLP:conf/emnlp/KarpukhinOMLWEC20} to combine the BM25 and DPR. The similarity score is computed as  $ f_{\phi}(X_{i},B_{j}) = sim(X_i,B_j) + \lambda~\text{BM25}(X_i,B_j)$, where $\lambda$ is a weight  to balance the two retrievers and was empirically set to 1 in our experiment. Based on this combined similarity score, we select the top-1 relevant bug-fix pair $(B_j,F_j)$ as a fix pattern to guide the patch generator for bug fixing. The hybrid retriever is expected to be more robust compared to retrievers that rely only on  either  lexical or semantic information.

\subsection{Retrieval-Augmented Patch Generator}\label{model:generator}
As shown in ~\cref{fig:racodet5_model}~{2}, given a buggy patch $X_i$, we search for a top relevant fix pattern $(B_j,F_j)$ and pass it to the patch generator to generate a fixed code patch $Y_i$.
We adopt a simple yet effective strategy to augment $X_i$ into $\hat{X}_i = X_i \oplus B_j \oplus F_j$  via appending the retrieved bug-fix pair into the source buggy patch.
Note that the patch generator module can be any sequence generation model.
Different from prior studies that directly adopt a generator optimized on natural language~\cite{DBLP:conf/icml/BerabiHRV21}, we propose to employ CodeT5, a code-aware  programming language model optimized for code.

\paragraph{\textbf{Training}}
We prepare the retrieval-augmented input to CodeT5 patch generator as $\hat{X}_i$ = ``\texttt{[CLS]} X$_i$ \texttt{[BUG]} $B_j$ \texttt{[FIX]} $F_j$'', where \texttt{[BUG]} and \texttt{[FIX]} are special tokens to separate the retrieved bug-fix pair from the  buggy patch. 
CodeT5's encoder takes  $\hat{X}_i$ as input  and  emits the fixed patch $Y_i$ from its decoder in an autoregressive manner (see \cref{model:task}). We consider two settings of the buggy patch $X_i$ where it
may or may not contain bug localization information. If it contains error information like error type, error message, and error line, the buggy patch will be augmented to ``error information \texttt{[SEP]} $X_i$'' to incorporate error information to help fix the bugs.
To train the patch generator, we adopt \emph{teacher forcing}~\cite{DBLP:journals/nn/ToomarianB92} to minimize the cross entropy loss $\mathcal{L}_{ce}$ over all training instances defined as: 
\begin{displaymath}\label{eq:lm}
    \mathcal{L}_{ce} = -\sum_{i=1}^{|\mathcal{D}|}\log(P_{\theta}(Y_i|\hat{X}_i))
\end{displaymath}

In teacher forcing, the decoder uses ground-truth context for faster convergence. We use the training set as the search codebase following~\cite{DBLP:conf/emnlp/ParvezACRC21}. To avoid information leakage, we do not allow the retriever to access the ground-truth bug-fix pair, otherwise the training loss would easily drop close to 0 as the generator can directly copy the retrieved fix as the target output.  This strategy  makes the training and evaluation process more compatible as the evaluation sets  are  not  overlapped with the training set as well.

\paragraph{\textbf{Inference with Beam Search}}
During  inference, as shown in \cref{fig:racodet5_model}~{3}, we employ  beam search to generate a ranked list of  fixed patch candidates for an input buggy patch, where the number of predictions is determined by the beam size $\mathcal{B}$. Concretely, at each decoding timestep, the beam search selects the most $\mathcal{B}$ promising fix candidates with the highest probability using a best-first search strategy. The search process is terminated when an \texttt{[EOS]} token notifying the end of sentence is emitted. The top ranked fix patch will be examined for its correctness by comparing with ground-truth fix patches or by validating against test suites, or by manual verification by software developers.

\section{Experimental Design}

\subsection{Dataset}
We  evaluate RAP-Gen on three APR datasets, namely TFix~\cite{DBLP:conf/icml/BerabiHRV21} in JavaScript, Code Refinement~\cite{DBLP:journals/tosem/TufanoWBPWP19} and Defects4J (v1.2)~\cite{DBLP:conf/issta/JustJE14} in Java. 
All datasets are originally collected from open source GitHub commits but based on different criteria for bug identification, where TFix is based on diagnostics from a JavaScript static analyzer, Code Refinement is based on repair-related commit message, and Defects4J is based on running the test suites. 
We report their data statistics in Table~\ref{table:data_stats}.

\subsubsection{TFix}
TFix~\cite{DBLP:conf/icml/BerabiHRV21} is a large-scale program repair dataset comprising JavaScript code patch pairs curated from 5.5 million GitHub commits. It  includes 52  error types (see Table~\ref{table:error_type_breakdown}) detected by a static analyzer ESLint\footnote{\url{https://eslint.org/}}~\cite{DBLP:conf/kbse/TomasdottirAD17}. In addition to error types, it provides rich error annotations such as error message and localized error line so that there is no need for fault localization like prior work~\cite{DBLP:conf/icse/JiangL021,DBLP:conf/sigsoft/ZhuSXZY0Z21}.
To prepare the input sequence, as illustrated in ~\cref{fig:case_study}(a), we follow~\cite{DBLP:conf/icml/BerabiHRV21} to combine all error information together with the buggy code patch into a single piece of text as the following: 
\begin{align*}
  &\textbf{fix } \textbf{\{error type\} } \textbf{\{error message\} } \textbf{\{error context: } \\
  &\textbf{Code Line N-1 + Buggy Line N + Code Line N+1\}}
\end{align*}
where error context consists of the given localized error line and its two neighboring code lines to form a buggy code patch. For the target sequence, it is obtained by replacing the error line into a fixed line in the error context. 
During data processing, we observed a \emph{duplication issue}  inside each data split and between data splits. Specifically, there are 114, 2, and 4 duplicates in the train, validation, and test split respectively, and  28, 34, and 4 duplicates for inter-split duplicates between train and test, train and test, validation and test splits respectively.
We filtered all these 243 duplicates  to get a deduplicated version of TFix  as shown in Table~\ref{table:data_stats}.

\begin{table}[!t]
\centering
\caption{Statistics of three program repair benchmarks.
}\label{table:data_stats}
\vspace{-0.5em}
\resizebox{\linewidth}{!}{
\begin{tabular}{llccc}
\toprule
 Benchmark & Version & Train  & Valid & Test   \\
\midrule
TFix& Original &84,846 & 9,454 & 10,504 \\ 
TFix& Deduplicated & 84,673 & 9,423 & 10,465  \\
Code Refinement& Small &46,680 & 5,835 & 5,835  \\
Code Refinement & Medium  & 52,364 & 6,545 & 6,545  \\
Defects4J        & v1.2 & - & - & 388  \\
Defects4J        & v2.0 & - & - & 430  \\
\bottomrule
\end{tabular}
}
\vspace{-0.5em}
\end{table}

\emph{Baseline Models.}
We compare  RAP-Gen with existing DL-based APR models including SequenceR~\cite{DBLP:journals/tse/ChenKTPPM21} and CoCoNuT~\cite{DBLP:conf/issta/LutellierPPLW020}. Besides, we compare a large pretrained model T5-large~\cite{DBLP:journals/jmlr/RaffelSRLNMZLL20}  which has been finetuned on TFix  to achieve the SoTA performance by~\cite{DBLP:conf/icml/BerabiHRV21}. 

\emph{Evaluation Metrics.}
We report  \textbf{Exact Match} (EM) accuracy and  \textbf{BLEU-4} score to evaluate program repair performance following~\cite{DBLP:conf/emnlp/0034WJH21} on  TFix. BLEU-4 is a looser metric to evaluate the degree of subword overlapping while EM is a more strict metric requiring the prediction to be identical to the ground-truth patch in a real commit. As a buggy program might have different ways to repair, we further employ \textbf{Error Removal} metric following \cite{DBLP:conf/icml/BerabiHRV21} to take various forms of fixes into account. The prediction is counted as correct for Error Removal if the existing error is removed and no new errors (detected by the static analyzer ESLint) is introduced after the fix. For all metrics, we present their results on a scale of 0-100 (\%) and a higher score represents better performance.

\subsubsection{Code Refinement}
Code Refinement~\cite{DBLP:journals/tosem/TufanoWBPWP19} contains bug-fix pairs at the function level, which are originally collected from public GitHub Archive\footnote{\url{https://www.gharchive.org/}} between March 2011 and
October 2017. They use Google BigQuery APIs to identify all Java commits having a message containing the patterns: (``fix'' or ``solve'') and (``bug'' or ``issue'' or ``problem'' or ``error'') to ensure the quality of the collected bug-fix function pairs. 
They normalized the functions via obfuscating identifiers with indexed tokens such as TYPE1, VAR1, METHOD1, etc. One data example can be found in ~\cref{fig:case_study}~(b). The dataset contains two data subsets which are determined by the number of tokens, i.e., \# of code tokens $\leq$ 50 for the small set and 50 $<$ \# of code tokens $\leq$ 100 for the medium set. Since the bug localization is not provided, the entire code fragment is taken as the source input sequence of our model. The target sequence is the refined version of the whole code snippet.  

\emph{Baseline and Metrics.} 
We compare our RAP-Gen with pretrained programming language models based on Transformers~\cite{DBLP:conf/nips/VaswaniSPUJGKP17}. One group of these models is the encoder-only models such as RoBERTa (code), CodeBERT~\cite{DBLP:conf/emnlp/FengGTDFGS0LJZ20}, and GraphCodeBERT~\cite{DBLP:conf/iclr/GuoRLFT0ZDSFTDC21}.
These encoder-only models require a randomly initialized decoder to generate the fix. Besides, we compare with encoder-decoder Transformer models  such as PLBART~\cite{DBLP:conf/naacl/AhmadCRC21} and CoTexT~\cite{DBLP:journals/corr/abs-2105-08645}.
NSEdit~\cite{DBLP:journals/corr/abs-2204-06643} is a language  model with encoder and decoder initialized from CodeBERT and CodeGPT~\cite{DBLP:journals/corr/abs-2102-04664}  respectively. It is finetuned to generate the fix via a neural-symbolic editing sequence and ranks as the current SoTA model on Code Refinement. 
We follow~\cite{DBLP:conf/emnlp/0034WJH21} to apply BLEU-4 and Exact Match to evaluate the Code Refinement datasets.

\subsubsection{Defects4J} 
Defects4J~\cite{DBLP:conf/issta/JustJE14} has been one of the most widely adopted APR benchmarks, which contains 835 real bug-fix patches in 17 open source GitHub projects. Each bug-fix example is accompanied with test cases to validate the fix. One example of Defects4J bugs can be found in~\cref{fig:case_study}(c), where ``-'' denotes a buggy line to be fixed and ``+'' represents the correct fix committed from a developer. A buggy line and its corresponding code context are combined to form the source input sequence while the target sequence is the fixed line. As Defects4J only has the test set, we use the project-specific training data curated by SelfAPR~\cite{SelfAPR} using self-supervised learning methods. Specifically, ~\citet{SelfAPR} proposes 16 perturbation rules on the correct past version of Defects4J to construct 1,039,873 synthetic bug-fix Java patches.
We use a subset of 830,240 training data that is available online.\footnote{\url{https://github.com/ASSERT-KTH/SelfAPR/tree/main/dataset}} 
For testing, we follow their exact settings to evaluate our models on 818 bugs from both Defects4J v1.2 and v2.0 (\cref{table:data_stats}), which covers both settings with ground-truth fault localization (perfect FL) and with predicted FLs from spectrum-based FL tools such as Gzoltar~\cite{GZoltar}.

\emph{Baselines and Metrics.} We compare  RAP-Gen with a broad set of SoTA DL-based APR models including SequenceR~\cite{DBLP:journals/tse/ChenKTPPM21}, 
CoCoNuT~\cite{DBLP:conf/issta/LutellierPPLW020},
CURE~\cite{DBLP:conf/icse/JiangL021},
RewardRepair~\cite{DBLP:conf/icse/YeMM22},
Recoder~\cite{DBLP:conf/sigsoft/ZhuSXZY0Z21},
DLFix~\cite{DBLP:conf/icse/Li0N20}, 
DEAR~\cite{DBLP:conf/icse/Li0N22},
BugLab~\cite{DBLP:conf/nips/AllamanisJB21},
and SelfAPR~\cite{SelfAPR}. 
For evaluation, we compute how many bugs can be correctly fixed on Defects4J based on unit testing and manual verification following prior work. We first run test suites to automatically identify plausible correct patches for each bug, followed by manual checking to completely verify its correctness. The correct predictions from our RAP-Gen are included in our artifact. For results of baselines, we cite the results of DLFix and DEAR from DEAR \cite{DBLP:conf/icse/Li0N22}, and  other  results from SelfAPR~\cite{SelfAPR}.

\begin{table}[!t]
\centering
\caption{Comparison results of CodeT5 on the original TFix. }\label{table:tfix-codet5}
\vspace{-0.5em}
\begin{tabular}{lcccc}
\toprule
 \multirow{2}{*}{Model} & \multicolumn{2}{c}{EM w/ spaces} & \multicolumn{2}{c}{EM w/o spaces}  \\
 \cmidrule(lr){2-3}\cmidrule(lr){4-5}
& Avg.      & W. Avg.     & Avg.      & W. Avg.  \\ 
\midrule
Naive Copy&  0.00 & 0.00 & 0.00& 0.00\\
SequenceR & 17.90 & - & -& - \\
CoCoNuT &11.70 & - & -& - \\

T5-small  &44.46 & 44.44&  44.52 & 44.60 \\
T5-base   &48.54 &47.63 & 48.72 & 47.70  \\
T5-large    &49.33 &\textbf{49.65} & 49.35 &49.70  \\

\midrule
CodeT5-small    &47.14&46.35&50.22& 50.31\\
~~- error-information &  45.97&45.80&49.26&49.70\\

CodeT5-base   &\textbf{50.88} & 49.42 &\textbf{54.30}& \textbf{53.57}\\
~~- error-information & 46.88&47.17 &50.36&51.25\\

\bottomrule
\end{tabular}
\vspace{-0.5em}
\end{table}
\subsection{Implementation Details} 
We adopt CodeT5-base~\cite{DBLP:conf/emnlp/0034WJH21} that contains 12 encoder layers and 12 decoder layers with the parameter size of 220M for   RAP-Gen.
We implement RAP-Gen using PyTorch and train it with AdamW~\cite{DBLP:conf/iclr/LoshchilovH19} optimizer. 
For the training of its neural components,
we run these experiments with NVIDIA A100-40G GPUs on the Google Cloud Platform.
For each benchmark, we finetune a DPR retriever  for 50 epochs 
using the contrastive loss $\mathcal{L}_{\text{infoNCE}}$ using a batch size of 64 and a learning rate of 2e-5.
We  finetune RAP-Gen generator for 30 epochs using a sequence generation loss $\mathcal{L}_{ce}$ using a batch size of 32 with a learning rate of 5e-5.
These best settings are obtained through a  grid search for hyper-parameter tuning: batch size in (16, 32, 64) and learning rate in (1e-4, 5e-5, 2e-5).
The training time of DPR retriever is 5-9 hours depending on the training size of the dataset, and the training time of RAP-Gen generator is within 2 days.  
For lexical-based retrievers, we use an open-sourced Python library\footnote{{https://pypi.org/project/rank-bm25}} of BM25, which can be efficiently trained on CPU within one hour with multi-processing.
During inference, we employ beam search with a beam size of 5 for the TFix and Code Refinement, and 100  for the Defects4J.


\begin{table}[!t]
\centering
\caption{Performance of RAP-Gen on the deduplicated TFix.  Results of T5-large and CodeT5-base are different from~\cref{table:tfix-codet5} due to the deduplication. 
}\label{table:tfix-raapr}
\vspace{-0.5em}
\begin{tabular}{lcc}
\toprule
  Model & EM & BLEU-4    \\
\midrule
T5-large (TFix)  & 49.58 & 76.96 \\
CodeT5-base  & 53.46& 78.92 \\
\midrule
RAP-Gen  & \textbf{54.15} & \textbf{79.66} \\
\bottomrule
\end{tabular}
\vspace{-0.5em}
\end{table}

\subsection{Research Questions}
To investigate the effectiveness of  RAP-Gen  on APR tasks, we seek to answer the following research questions (RQs):

\noindent\textbf{RQ1: Comparative study with DL-based APR models on TFix.} How does RAP-Gen perform to repair JavaScript linter-flagged coding errors on TFix compared with other DL-based APR approaches?

\noindent\textbf{RQ2: Analysis of RAP-Gen predictions on TFix.}
How does RAP-Gen repair TFix bugs for different error types and patch lengths? What fix operations do RAP-Gen adopt in repairing bugs?

\noindent\textbf{RQ3: Comparative study with DL-based APR models on Code Refinement.}
How does RAP-Gen perform to repair Java commit-related bugs compared with other DL-based APR approaches?

\noindent\textbf{RQ4: Analysis of our hybrid patch retriever.} Can our hybrid patch retriever find relevant fix pattern to guide APR?

\noindent\textbf{RQ5: Comparative study with DL-based APR models on  Defects4J?}
How does RAP-Gen perform to repair  Java bugs in open source projects compared with other DL-based APR approaches?

\section{EXPERIMENTAL RESULT}

\subsection{RQ1: Comparative study with DL-based APR models on TFix} \label{rq1}

\subsubsection{Improved TFix Evaluation} 

The original TFix benchmark employs the direct average of exact match (EM) accuracy across 52 error types as the main  metric. However, as shown in the Table~\ref{table:error_type_breakdown}, these error types have a rather imbalanced distribution, e.g., the major error type ``no-invalid-this''  has 16,166 instances while the least error type ``no-new-symbol'' has only 10 instances. As such, it is more reasonable to employ the weighted average to take the error type distribution into account. Besides,  we spot another limitation of its exact match evaluation  that if the predicted fix contains one more whitespace such as a space or new line than the ground-truth fix, it would be regarded as a wrong exact match. However,  extra whitespaces do not impact the correctness for JavaScript programs. Therefore, we propose to use the weighted average of EM w/o spaces, which normalizes the whitespaces before computing the EM to exclude the effects of the mismatch in whitespaces. As we find there is a  duplication issue in the TFix dataset, we  also report the results on its deduplicated version.

\begin{table}[!t]
\centering
\caption{Performance  breakdown on 52 error types on TFix.
}\label{table:error_type_breakdown}
\vspace{-0.5em}
\resizebox{1\linewidth}{!}{%
\begin{tabular}{l   c c c | l   c c c}
\toprule
Error Type & \#Samples & T5-large & RAP-Gen & Error Type & \#Samples & T5-large & RAP-Gen \\
 \midrule
no-new-symbol         & 10                    & 100.00                & 100.00                & no-extra-bind          & 674                   & 70.59                 & 73.53                 \\
no-compare-neg-zero   & 13                    & 0.00                  & 0.00                  & no-case-declarations   & 723                   & 58.90                 & 67.12                 \\
no-ex-assign          & 40                    & 25.00                 & 25.00                 & no-fallthrough         & 743                   & 76.00                 & 77.33                 \\
for-direction         & 50                    & 40.00                 & 60.00                 & no-inner-declarations  & 830                   & 38.10                 & 46.43                 \\
no-unsafe-finally     & 63                    & 42.86                 & 14.29                 & no-array-constructor   & 980                   & 86.73                 & 85.71                 \\
use-isnan             & 71                    & 37.50                 & 25.00                 & no-constant-condition  & 1,251                 & 48.78                 & 54.47                 \\
no-class-assign       & 111                   & 41.67                 & 50.00                 & generator-star-spacing & 1,396                 & 67.86                 & 72.86                 \\
no-dupe-class-members & 117                   & 8.33                  & 8.33                  & no-extra-boolean-cast  & 1,458                 & 54.11                 & 58.22                 \\
no-func-assign        & 147                   & 46.67                 & 60.00                 & no-cond-assign         & 1,472                 & 45.21                 & 47.95                 \\
no-empty-pattern      & 178                   & 27.78                 & 44.44                 & no-process-exit        & 1,514                 & 32.89                 & 37.50                 \\
no-unused-labels      & 187                   & 52.63                 & 63.16                 & no-empty               & 2,055                 & 26.70                 & 31.55                 \\
no-duplicate-case     & 195                   & 65.00                 & 60.00                 & no-dupe-keys           & 2,181                 & 53.42                 & 55.25                 \\
getter-return         & 203                   & 52.38                 & 61.90                 & prefer-spread          & 2,466                 & 45.08                 & 45.49                 \\
no-sparse-arrays      & 237                   & 25.00                 & 45.83                 & no-useless-escape      & 2,920                 & 35.15                 & 40.61                 \\
no-const-assign       & 277                   & 35.71                 & 42.86                 & no-console             & 3,067                 & 73.62                 & 73.94                 \\
no-global-assign      & 318                   & 59.38                 & 68.75                 & guard-for-in           & 3,231                 & 41.98                 & 45.37                 \\
no-new-wrappers       & 360                   & 27.78                 & 38.89                 & no-throw-literal       & 4,075                 & 72.06                 & 74.51                 \\
no-this-before-super  & 413                   & 47.62                 & 69.05                 & no-debugger            & 4,164                 & 94.48                 & 94.24                 \\
no-unsafe-negation    & 423                   & 72.09                 & 76.74                 & prefer-rest-params     & 4,534                 & 35.68                 & 43.61                 \\
require-yield         & 429                   & 72.09                 & 76.74                 & no-unreachable         & 4,725                 & 63.85                 & 64.69                 \\
no-extend-native      & 443                   & 31.11                 & 26.67                 & no-extra-semi          & 5,969                 & 82.61                 & 83.61                 \\
no-new-object         & 446                   & 71.11                 & 66.67                 & no-redeclare           & 6,381                 & 49.45                 & 59.78                 \\
no-caller             & 446                   & 20.00                 & 22.22                 & comma-style            & 6,395                 & 46.48                 & 52.11                 \\
constructor-super     & 464                   & 59.57                 & 70.21                 & no-unused-vars         & 7,765                 & 51.87                 & 56.11                 \\
valid-typeof          & 539                   & 51.85                 & 51.85                 & no-undef               & 10,636                & 22.65                 & 27.35                 \\
no-self-assign        & 610                   & 34.43                 & 44.26                 & no-invalid-this        & 16,166                & 37.48                 & 44.13                \\
 \midrule
&&&& Sum/W. Avg. &  104,561      &  49.58          & \textbf{54.15}\\

\bottomrule
\end{tabular}
}
\vspace{-0.5em}
\end{table}
\begin{figure}[!t]
  \centering
  \vspace{-0.5em}
  \includegraphics[width=0.8\linewidth]{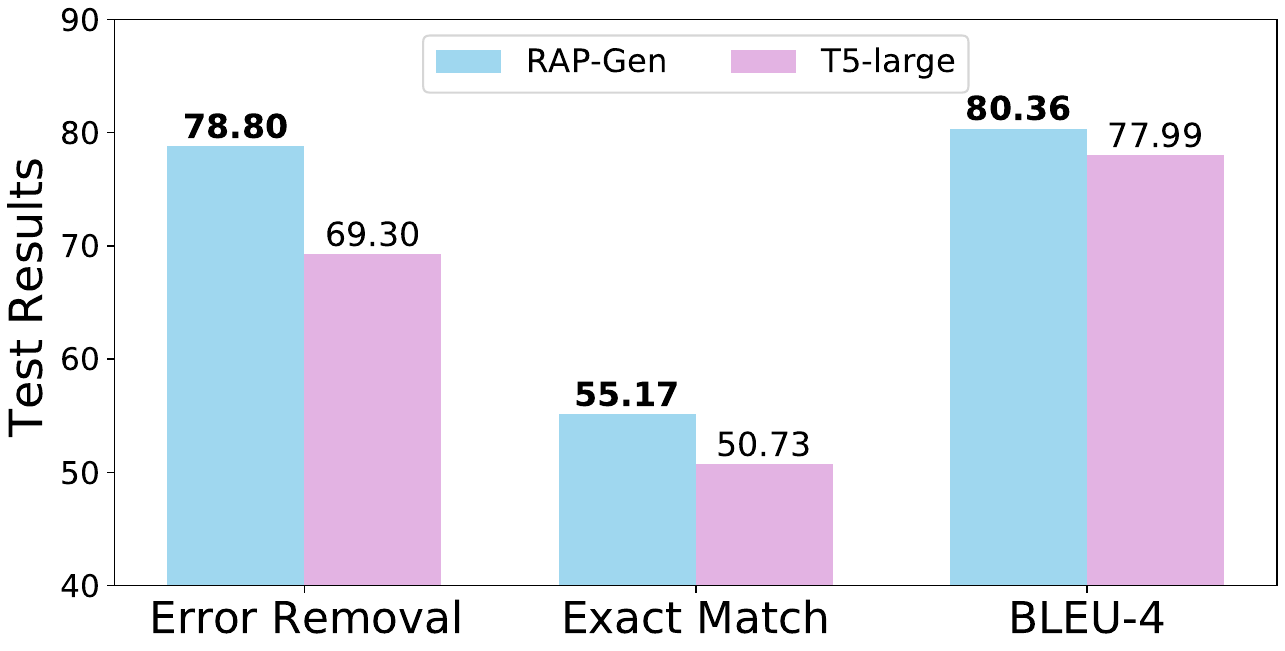}
  \vspace{-0.5em}
  \caption{Error removal comparison on TFix, where error removal is well aligned with exact match and BLEU-4 scores. }
  \label{fig:error-removal}
  \vspace{-0.5em}
\end{figure}
\subsubsection{CodeT5 Results} 
We compare CodeT5 models with other DL-based baselines on  TFix  and show results in Table~\ref{table:tfix-codet5}. 
For the original metric of average EM w/ spaces, CodeT5-base (50.88) also yields a better accuracy than T5-large (49.33), given that it has much larger model size ($\sim3.5\times$ of CodeT5-base: 770M vs. 220M).  If we focus on a more reasonable  average EM w/o spaces, CodeT5-base significantly boost the performance, with around 5 absolute accuracy improvement (49.35$\rightarrow$54.30) over T5-large.  Based on the weighted average EM w/o spaces, both CodeT5-small (50.31) and CodeT5-base (53.57) outperform all the baselines including T5-large (49.70). This shows CodeT5 models with code-aware pretraining on large-scale source code  have a better understanding of program. For TFix evaluation, we employ EM to denote the weighted average EM w/o spaces.
We  perform an ablation study to remove the error information including error type and error message from the input sequence, where we observe both CodeT5-small and CodeT5-base models have a consistent performance downgrade, revealing that it is helpful to inform  which types of error they need to fix for APR models.

\subsubsection{RAP-Gen Results}
We report the results of our RAP-Gen model on the deduplicated TFix benchmark in Table~\ref{table:tfix-raapr}, where the results are slightly different due to data size changes after duplication.  
Results show that RAP-Gen significantly outperforms  T5-large  (49.58$\rightarrow$54.15 EM).
This indicates retrieval-augmented generation is a viable and effective approach for APR and both semantic information  and lexical information are crucial to retrieve relevant fix patterns.
We present one case in \cref{fig:case_study}~(a), where we can observe RAP-gen successfully repairs the bug with the guidance of retrieved fix pattern  while CodeT5 without retrieval gives a wrong fix.

\begin{table}[!t]
\centering
\caption{Analysis of error line removal operation on TFix.
}\label{table:wrong_error_line_removal}
\begin{tabular}{l  c c c }
\toprule
 & T5-large  &  CodeT5 &  RAP-Gen  \\
\midrule

\# Ground-truth EL Removal & 2,381          & 2,381 & 2,381          \\
\# Predicted EL Removal    & 1,882       & 1,925 & 1,922          \\
\# Correct EL Removal      & 1,811        & 1,858 & 1,866          \\
\# False Positive          & 71               & 67    & 56             \\
\midrule

Precision (\%)             & 96.23           & 96.52 & \textbf{97.09} \\
Recall (\%)                & 76.06 &  78.03 & \textbf{78.37}          \\
F1 (\%)                    & 84.96 & 86.30 & \textbf{86.73}  \\
\bottomrule
\end{tabular}
\end{table}

\begin{figure}[!t]
  \centering
  \includegraphics[width=1\linewidth]{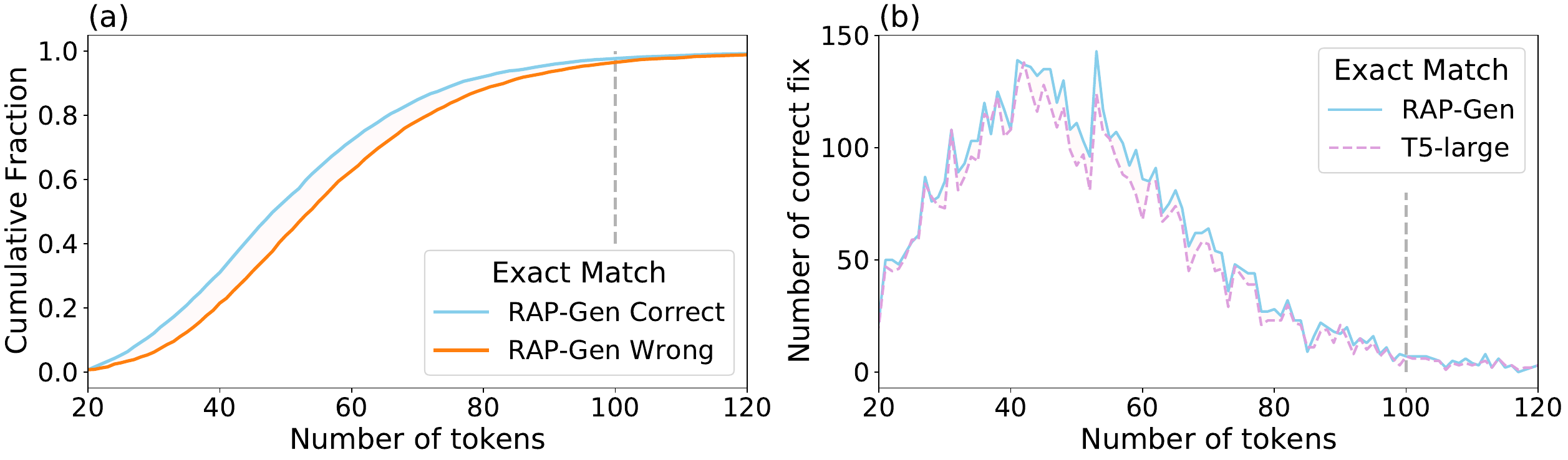}
  \caption{(a): cumulative fraction of programs by number of tokens in  the source buggy patch, grouped by whether RAP-Gen can accurately fix.  (b): distribution of correct fix over number of tokens for RAP-Gen and T5-large. }
  \label{fig:patch-length}
\end{figure}

\subsubsection{Error Removal Evaluation}\label{sec:error_removal_tfix}
Though exact match can  ensure correctness of machine-generated patches, it might be a too strict metric to consider other forms of correct fixes. 
Therefore, we follow~\cite{DBLP:conf/icml/BerabiHRV21} to employ the error removal metric, where a fix is counted as correct if the error is removed and no new error is introduced. The error detection is based on a static analyzer ESLint.
We report  error removal together  with EM and BLEU-4 results on a large subset of 6,793 instances\footnote{Some source files are unavailable to reproduce this metric on the full test set.} in ~\cref{fig:error-removal}.  We observe that RAP-Gen significantly improves  error removal accuracy over T5-large (69.30$\rightarrow$78.80). The larger gain compared to EM and BLEU-4  implies that RAP-Gen is more capable of producing various forms of good fixes. Additionally, EM is well aligned with the looser metric of error removal.

\subsection{RQ2: Analysis of RAP-Gen  on TFix} \label{RQ3}

\subsubsection{Performance Breakdown on  Error Types}
We list the performance breakdown for 52 error types on the deduplicated TFix in Table~\ref{table:error_type_breakdown}. RAP-Gen outperforms the previous SoTA T5-large in 40/52 error types. Especially for the major error type ``no-invalid-this'', RAP-Gen improves its exact match from T5-large's 37.48 to 44.13, i.e. repairing more 107 instances.  
In total, RAP-Gen correctly repairs more 478 bugs than T5-large with a much smaller model size.

\begin{table}[!t]
\centering
\caption{Performance of RAP-Gen on the Code Refinement.}\label{table:refine-raapr}
\begin{tabular}{lcccc} 

\toprule
\multirow{2}{*}{Model} & \multicolumn{2}{c}{Small} & \multicolumn{2}{c}{Medium}\\ 
\cmidrule(lr){2-3}\cmidrule(lr){4-5}
  & EM & BLEU-4 & EM & BLEU-4 \\ 
 
\midrule
Naive Copy & 0.00 & 78.06 & 0.00 & 90.91 \\
LSTM&10.00&76.76&2.50&72.08\\
Transformer&14.70&77.21&3.70&89.25\\
RoBERTa (code)&15.90&77.30&4.10&90.07\\
CodeBERT&16.40&77.42&5.16&91.07\\
GraphCodeBERT&17.30&80.02&9.10&91.31\\
PLBART&19.21&77.02&8.98&88.50\\
CoTexT&21.58&77.28&13.11&88.40\\
NSEdit&24.04&71.06&13.87&85.72\\
CodeT5 & 21.61& 77.43&13.96&87.64 \\
\midrule
RAP-Gen & \textbf{24.80}&78.28&\textbf{15.84}&90.01  \\
\bottomrule
\end{tabular}
\end{table}

\subsubsection{Fix Operation Analysis}
We analyze what  fix patterns are performed by our models on TFix. 
We observe a large proportion of fix consists of deletion operations compared to the code insertion and replacement operations. 
We find that fix operations consist of code insertion (12.5\%), replacement (8.1\%), deletion (47.9\%), insertion and replacement (6.9\%), insertion and deletion (8.2\%), replacement and deletion (7.2\%), and all three manners (9.2\%). Earlier studies~\cite{DBLP:conf/issta/QiLAR15,DBLP:conf/sigsoft/TanYPR16} also reflect that the deletion operation is one of the most common fix patterns.
Besides, we find one dominating  fix operation is  error line (EL) removal, which is to simply remove the error line from the buggy code and accounts for around 23\% in the test set. 
We show how models perform this operation in Table~\ref{table:wrong_error_line_removal}. We observe RAP-Gen achieves the best precision, recall, and F1 scores with a lowest false positive count of 56 compared to CodeT5's 67 and T5-large's 71. This indicates that RAP-Gen is able to learn more diverse bug fix patterns instead of over relying on the trivial error line removal pattern.

\subsubsection{Patch Length Analysis}
We analyze the impacts of  patch length \cref{fig:patch-length}.
\cref{fig:patch-length} (a) shows the cumulative fraction of
buggy patches by its patch length grouped based on their outcome.
We find  the patches successfully
repaired by RAP-Gen tend to be shorter than those where it
fails. 
\cref{fig:patch-length} (b) shows the distribution of correct fixes by its buggy patch length, where RAP-Gen can  repair more bugs than T5-large especially for patches  with 40 to 60 tokens.

\subsection{RQ3: Comparative study with DL-based APR models on Code
Refinement} 
We report the comparison results on Code Refinement  in Table~\ref{table:refine-raapr}. All baseline results  are directly obtained from their original papers. 
We first observe that  ``Naive Copy'' gives a pretty high BLEU-4 score but with a zero exact match, indicating the buggy code and its fix has a large overlap and exact match should be employed as the primary  metric. 
Among the baselines, NSEdit is a very competitive one with a best result (24.04 EM) on the small subset and CodeT5 gives the best result (13.96 EM) on the medium set.
The lower results on the medium set compared to the  small set indicates that longer buggy functions are more difficult to  fix, which is aligned with observations in \cref{fig:patch-length} (a).
Overall, RAP-Gen achieves new SoTA results on two subsets  with 24.80 EM for small set and 15.84 EM for medium set.
This again confirms that retrieved fix patterns provide  helpful signals to guide the program repair and the hybrid retriever is more robust by using both lexical and semantic information. 
\cref{fig:case_study} (b) shows one case where the retrieved fix pattern (error line removal) helps RAP-Gen to successfully fix the bug.

\begin{table}[!t]
\centering
\caption{Effects of retriever modules in RAP-Gen.}\label{table:retriever_module}
\resizebox{1\linewidth}{!}{%
\begin{tabular}{lcccc} 

\toprule
Retriever  & TFix & Refine-Small & Refine-Medium\\ 
 \midrule
No Retriever & 53.46 &21.61 & 13.96 \\
 \midrule
Random     & 52.98 & 21.25 & 13.53  \\
BM25       & 53.88 & 23.82 & 15.37  \\
CodeBERT    & 52.96 & 22.28 & 15.42 \\
CodeT5  & 53.93 & 24.37 & 15.60\\
Hybrid (BM25+CodeT5) & 54.15 & 24.80 & 15.84 \\
\bottomrule
\end{tabular}
}
\end{table}

\subsection{RQ4: Analysis of  Hybrid Patch Retriever}\label{RQ4}

We investigate how different retrieval modules affect the APR performance in the retrieval-augmented generation setting  in~\cref{table:retriever_module}.
We first compare with a Random baseline via randomly retrieving bug-fix pairs from the codebase. 
The consistent performance downgrade compared to ``no retriever'' implies that randomly retrieved fix patterns cannot provide useful guiding signals for APR.  
Then we compare our hybrid retriever in RAP-Gen  with different retrievers:  sparse BM25 retrievers, and dense retrievers based on  CodeBERT or CodeT5.
We observe that CodeT5-based  retrievers outperforms either BM25 or CodeBERT-based retrievers, while our hybrid retriever  combining both BM25 and CodeT5 achieves the best APR performance, validating the effectiveness of our retriever module design in RAP-Gen.

We further analyze the performance of our retrievers in terms of lexical and semantic matching between the query and the top retrieved patches. We employ the BLEU-4 score to measure their subtoken overlap for lexical matching, while for semantic matching, we compute the cosine similarity (CosSim) between their dense vectors encoded by our fine-tuned DPR retriever.  
Table~\ref{table:retriever_eval} shows the performance of our retrievers on both TFix and Code Refinement benchmarks. The first row indicates the lower-bound performance via randomly retrieving bug-fix pairs from the codebase, where we observe this Random baseline achieves much lower scores in both lexical and semantic matching. 

For lexical matching, BM25 outperforms  DPR (CodeT5-based) on TFix but underperforms on two Code Refinement subsets. We anticipate that it is due to the data difference between TFix and Code Refinement, where the latter employs obfuscated identifiers (e.g., VAR1, VAR2, ...) that hinders the performance of the lexical-based BM25 retriever. The hybrid retriever achieves the best lexical matching on all datasets, revealing  the semantic information can complement to the lexical information.
For semantic matching, DPR achieves the best results on all datasets, which is not surprising as it is optimized towards the identical objective. Notably, our hybrid retriever achieves slightly lower results than DPR but much better results than BM25, implying it can balance both lexical and semantic information and be more robust than the lexical-based retrievers, which are  sensitive to the choices of identifier naming.

\begin{table}[!t]
\centering
\caption{Lexical (BLEU-4) and semantic (CosSim) retrieval matching results on  TFix and Code Refinement benchmarks.}\label{table:retriever_eval}
\resizebox{1\linewidth}{!}{%
\begin{tabular}{lcccccc} 

\toprule
\multirow{2}{*}{Retriever} & \multicolumn{2}{c}{TFix} & \multicolumn{2}{c}{Refine-Small}
&\multicolumn{2}{c}{Refine-Medium}\\ 
\cmidrule(lr){2-3}\cmidrule(lr){4-5}\cmidrule(lr){6-7}
  & BLEU-4 & CosSim  & BLEU-4&  CosSim & BLEU-4 &  CosSim \\ 
 \midrule
 Random & 0.1& 35.5  & 14.6 & 35.4& 14.6& 30.6   \\
BM25 & 23.7& 70.9  & 41.5 & 68.5& 39.0& 66.6   \\
DPR & 21.7 & \textbf{75.4}   & 54.4 & \textbf{84.9}& 44.3 & \textbf{81.3} \\
Hybrid & \textbf{24.4}& 73.4 & \textbf{57.4} & 84.2   & \textbf{45.0}& 80.9\\
\bottomrule
\end{tabular}
}
\end{table}
\subsection{RQ5: Comparative study with DL-based APR models on Defects4J} \label{RQ5}

\subsubsection{RAP-Gen Results} 
We compare RAP-Gen with other SoTA DL-based APR baselines on Defects4J~\cite{DBLP:conf/issta/JustJE14} v1.2 and v2.0 in \cref{table:defects4j_eval}. We consider two settings with  spectrum-based fault localization (FL) and with the perfect FL. 
Note that all the baseline results are  cited from SelfAPR~\cite{SelfAPR} and DEAR~\cite{DBLP:conf/icse/Li0N22}.
For a fair comparison, we follow common practice to adopt 
the same 5-hour timeout, a beam size of 100, an ensemble strategy as Recoder~\cite{DBLP:conf/sigsoft/ZhuSXZY0Z21} for RAP-Gen.

As shown in \cref{table:defects4j_eval}, our RAP-Gen achieves new SoTA performance  under perfect FL by repairing the largest set of bugs (72 bugs in v1.2 and 53 bugs in v2.0) compared to other baselines. Particularly, it repairs 7 and 8 more bugs than the previous SoTA SelfAPR in v1.2 and v2.0 respectively.
For the results with spectrum-based FL, RAP-Gen achieves the second-best performance, which are very competitive to the SoTA models on both v1.2 (48 vs. Recoder's 49) and v2.0 (26 vs. SelfAPR's 28). Considering both v1.2 and v2.0 bugs, it repairs 74 bugs in total, surpassing either Recoder's 68 or SelfAPR's 67 bugs. 
Overall, both results with or without perfect FL validate the superiority of our RAP-Gen over other DL-based baselines.
Notably, compared to many of these models, our RAP-Gen exhibits another advantage of being a language agnostic model that can generalize to other APR use cases. By contrast, Recoder requires to learn edits over AST and SelfAPR requires the test execution diagnostics, making them inapplicable or limited to deal with fragmented code snippets that cannot be parsed into ASTs or other APR scenarios without test cases.

We investigate to what extent RAP-Gen can complement existing APR models, including Recoder~\cite{DBLP:conf/sigsoft/ZhuSXZY0Z21}, RewardRepair~\cite{DBLP:conf/icse/YeMM22}, and SelfAPR~\cite{SelfAPR}.
Compared with these SoTA DL-based APR approaches, RAP-Gen repairs 13 and 12 unique bugs for Defects4J v1.2 and v2.0 respectively, which are never correctly addressed by any other DL-based APR approaches, veifying that our RAP-Gen can complement to other top-performing APR approaches. 
We further show a case in \cref{fig:case_study} (c) and find that RAP-Gen successfully fixes the Chart-9 bug but in a different form with the developer's fix.

\begin{table}[!t]
\centering
\caption{Performance of RAP-Gen on Defects4J v1.2 and v2.0.}\label{table:defects4j_eval}
\resizebox{1\linewidth}{!}{%
\begin{tabular}{lcccc} 
\toprule
\multirow{2}{*}{Model} & \multicolumn{2}{c}{Spectrum-based FL} & \multicolumn{2}{c}{Perfect FL}\\ 
\cmidrule(lr){2-3}\cmidrule(lr){4-5}
  & v1.2 & v2.0 & v1.2 & v2.0 \\ 
\midrule
SequenceR~\cite{DBLP:journals/tse/ChenKTPPM21}     & -  & -  & 14  & -           \\
BugLab~\cite{DBLP:conf/nips/AllamanisJB21}       & -  & -  & 17 & 6           \\
DLFix~\cite{DBLP:conf/icse/Li0N20}          & 30 & -  & 40  & -       \\
CoCoNuT~\cite{DBLP:conf/issta/LutellierPPLW020}       & -  & -  & 43  & -           \\
RewardRepair~\cite{DBLP:conf/icse/YeMM22}  & 27 & 24 & 44  & 43          \\

DEAR~\cite{DBLP:conf/icse/Li0N22}          & 47 & -  & 53  & -       \\
CURE~\cite{DBLP:conf/icse/JiangL021}          & -  & -  & 55  & -           \\
Recoder~\cite{DBLP:conf/sigsoft/ZhuSXZY0Z21}       & \textbf{49} & 19 & 64  & -           \\

SelfAPR~\cite{SelfAPR}      & 39 & \textbf{28} & 65 & 45          \\
\midrule
CodeT5       & 27 & 13  & 58 & 28          \\
RAP-Gen      & 48 & 26  & \textbf{72} & \textbf{53}  \\
\bottomrule
\multicolumn{5}{l}{In the table cells, it represents the number of correct patches.}\\
\multicolumn{5}{l}{`-' indicates data unavailability.}\\
\end{tabular}
}
\vspace{-1em}
\end{table}
\subsubsection{Effects of Retrieval from Various Fix Patterns.}
\begin{figure}[!t]
  \centering
  \vspace{-0.5em}
  \includegraphics[width=1\linewidth]{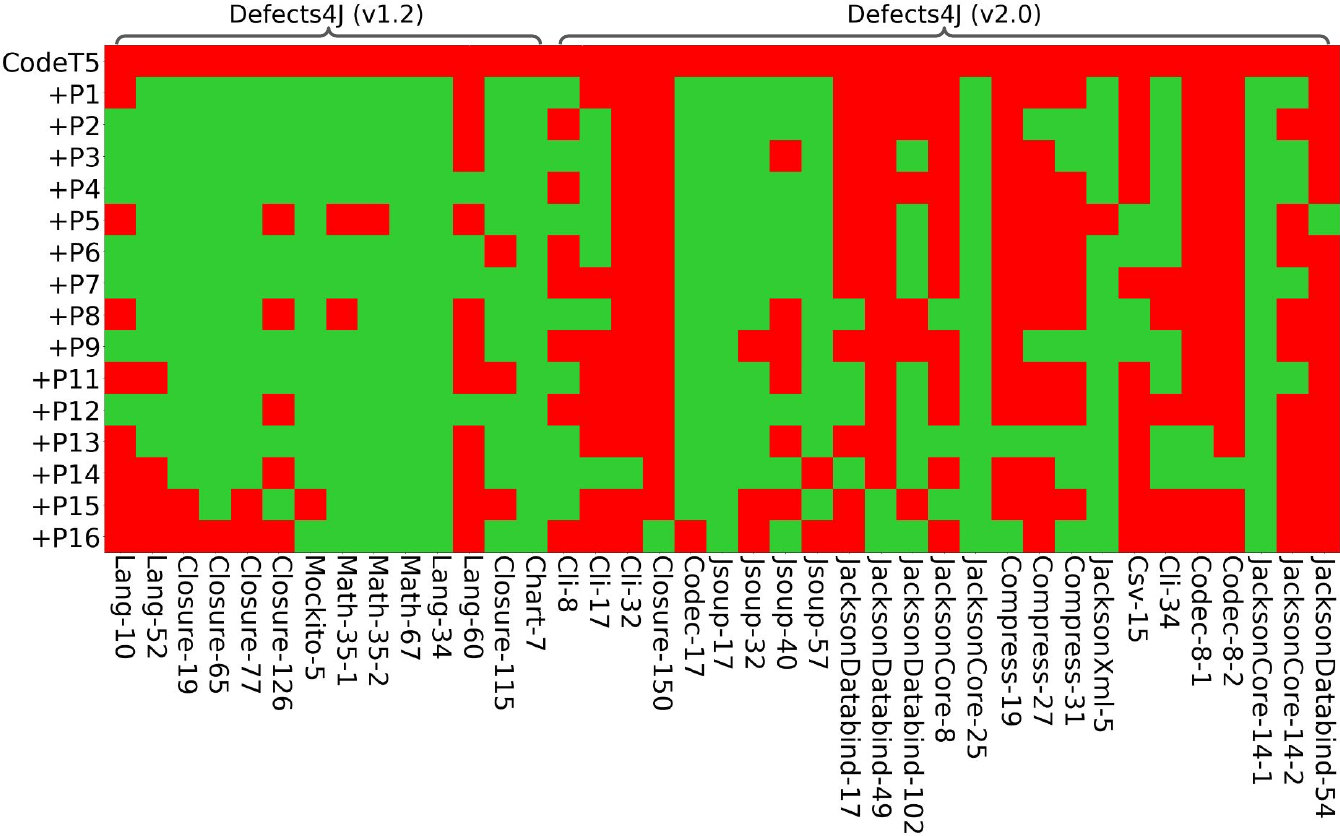}
  \caption{Effects of retrievals from various fix patterns over 39 bugs from Defects4J v1.2 and v2.0, which  RAP-Gen can fix  (green) and CodeT5 cannot fix (red). We represent each  bug on the X-axis and use the color to denote its fixing outcome under different retrieval schemes on the Y-axis.}
  \label{fig:pattern_analysis}
  \vspace{-0.5em}
\end{figure}

We analyze how retrieving a bug-fix sample from various fix patterns will affect the APR performance. For this analysis, as shown in \cref{fig:pattern_analysis}, we select 39 bugs from Defects4J v1.2 and v2.0 which CodeT5 cannot fix (red) and RAP-Gen can fix (green) under the setting with perfect FL.  For the categorization of fix patterns, we base on the 16 perturbation rules devised from SelfAPR~\cite{SelfAPR} and use its training set for each rule as a separate retrieval codebase. We retrieve the top-1 bug-fix sample from the codebase for each rule or fix pattern (denoted as P1 to P16) and examine whether it can improve CodeT5's performance after using the guiding signals from such retrieval in RAP-Gen.

From \cref{fig:pattern_analysis}, we observe that retrievals from various fix patterns in RAP-Gen are generally helpful in correcting CodeT5's predictions on Defects4J bugs. We find that most of bugs in v1.2 can be fixed after retrieval from many different patterns, while for v2.0, there are some bugs where only a few fix patterns are applicable, e.g., the P16 for Closure-150 and P5 for JacksonDatabind-54. Across various fix patterns, we find that the P13 of ``insert an existing block'' and P14  of ``delete statement'' are applicable to most bugs, indicating these are key fix patterns for repairing Defects4J bugs.

\section{Threats to validity}
\paragraph{Construct Validity} 
We evaluated  RAP-Gen on three APR benchmarks: TFix in JavaScript, Code Refinement and Defects4J in Java. On TFix, we spotted a duplication issue  and  removed 243 intra-split or inter-split duplicates out of total 104,804 data instances.  This might slightly impact the comparison between our model and the  TFix (T5-large) model. 
We mitigate this threat by reporting the results of our model on the original TFix dataset and also the results of TFix model on the deduplicated test set.
On Code Refinement, unlike the pairs in TFix can be validated by a static analyzer, its bug-fix pairs are curated from GitHub commits with a bug-fix related commit message, and only a  portion of them are manually verified~\cite{DBLP:journals/tosem/TufanoWBPWP19}.
There is a chance that some pairs are invalid (not related to the bug fix), which brings potential threats to the reliability of the evaluation on this dataset. 

\paragraph{Internal Validity}
The threats to internal validity mainly lie in the hyper-parameter search stage for RAP-Gen.
As a neural model, its performance is highly affected by the choice of hyper-parameters. To alleviate such threats, we conduct a grid search to tune a better set of hyper-parameters but we still cannot claim they are the best. 

\paragraph{External Validity}
We only evaluated our RAP-Gen model on JavaScript and Java programs and do not study its generalization to other programming languages (PLs). However, our approach is language-agnostic as we do not employ any code-specific features like ASTs and can be applied in a drop-in fashion to other PLs. Besides, our  evaluation on three APR datasets in two PLs should be comprehensive enough to verify the effectiveness of our approach.

\section{CONCLUSION}
We present a novel retrieval-augmented patch generation (RAP-Gen) framework for automatic program repair, a fundamental task in software engineering to reduce developers' manual efforts in debugging.  RAP-Gen consists of two components: a hybrid patch retriever to retrieve relevant fix patterns for a query buggy patch and a patch generator to synthesize the fixed patch based on both  buggy patch and its retrieved guiding fix patterns.
In addition, we propose to leverage a powerful code-aware pretrained language model CodeT5 as the backbone of RAP-Gen to facilitate both patch retrieval and generation in a unified manner. Comprehensive results on three diverse APR benchmarks in JavaScript and Java have demonstrated the effectiveness and superiority of our RAP-Gen model over  existing deep learning-based APR approaches.

\section{Data Availability}
 Our code and models can be found in this link (\url{https://figshare.com/s/a4e95baee01bba14bf4b}) to reproduce the results in this paper.
\bibliographystyle{ACM-Reference-Format}
\bibliography{fse23}

\end{document}